\documentclass[11pt]{article}
\usepackage{amsmath,amssymb,amsthm}
\usepackage[a4paper,top=3cm,bottom=3cm,left=3cm,right=3cm]{geometry}
\usepackage[ruled,vlined]{algorithm2e}
\usepackage{xcolor}

\usepackage{authblk}

\newtheorem{theorem}{Theorem}
\newtheorem{lemma}{Lemma}
\newtheorem{proposition}{Proposition}
\newtheorem{corollary}{Corollary}
\theoremstyle{definition}
\newtheorem{definition}{Definition}

\newcommand{\glorule}{\ensuremath{{\cal F}}}
\newcommand{\ZZ}{\mathbb{Z}}

\newcommand{\zzd}{{\mathbb{Z}^d}}
\newcommand{\Fp}{\mathbb{F}_p}
\newcommand{\LP}{\mathbb{L}_p}
\newcommand{\KP}{\mathbb{K}_p}

\newcommand{\Ker}{\mathrm{Ker}}

\newcommand{\Imma}{\mathrm{Im}}
\newcommand{\Cyl}{\mathrm{Cyl}}
\newcommand{\ie}{i.e.,\ }

\begin{document}

\title{Strong Transitivity of Group Cellular Automata: Characterization and Decidability}

\author[1]{Niccol\`o Castronuovo}
\author[2]{Alberto Dennunzio}
\author[1]{Luciano Margara}
\affil[1]{Department of Computer Science and Engineering, University of Bologna, Campus of Cesena, Italy}
\affil[2]{Department of Informatics, Systems and Communication, University of Milano-Bicocca, Italy}

\date{}
\maketitle

\begin{abstract}
A discrete-time dynamical system is strongly transitive if every point can be reached, in finitely many steps, from every non-empty open set. We study this property for group cellular automata, \ie cellular automata over a finite group alphabet whose global map is a group endomorphism, in every dimension $d\geq1$. We prove that no group cellular automaton over a non-abelian group is strongly transitive, and we develop a decomposition of group cellular automata over abelian groups, via quotients by power subgroups, which preserves strong transitivity. For linear cellular automata over $(\mathbb Z/p\mathbb Z)^n$ we use Pontryagin duality to characterize strong transitivity in terms of the characteristic polynomial $\chi_A$ of the associated matrix $A$ of Laurent polynomials: the automaton is strongly transitive if and only if $\det(A)\neq0$ and no irreducible factor of $\chi_A$ has an invertible constant term. 
Combining these results, we obtain that strong transitivity is decidable for group cellular automata in every dimension. For every fixed dimension $d$, we give an explicit decision procedure whose running time is polynomial in the group size $N$, the number $k$ of neighbours, and 
the neighbourhood radius $\varrho$.

\end{abstract}

\section{Introduction}\label{sec:intro}
 
Cellular automata (CAs) are discrete-time dynamical systems defined by the synchronous application of a local rule on a regular lattice $\zzd$ of cells, each of which takes values in a finite alphabet. Despite their simple definition, they exhibit a wide range of dynamical behaviors, and many of their dynamical properties are undecidable in general~\cite{Kari2005survey}. This motivates the search for natural classes of CAs in which such properties become decidable and admit explicit characterizations.
 
A continuous map $f$ of a compact metric space $X$ is \emph{strongly transitive} if $\bigcup_{n\geq0}f^n(U)=X$ for every non-empty open set $U\subseteq X$, that is, if every point of $X$ can be reached from every open set. Strong transitivity is a strong form of topological transitivity: a strongly transitive map is surjective, and the backward orbit of every point is dense. For CAs, strong transitivity is closely related to expansivity: every positively expansive one-dimensional CA is strongly transitive~\cite{BlanchardCervelleFormenti2005}, while in dimension $d\geq2$ there are no positively expansive CAs~\cite{Kari2005survey}. 

Strong transitivity has been investigated for general CAs~\cite{AcerbiDF07,AcerbiDF09,BlanchardCervelleFormenti2005,DennunzioF08DLT,DennunzioFW14} and, in particular, for linear and additive CAs and, more generally, for GCAs (including linear and additive CA)~\cite{CattaneoDM04,DennunzioFGM19,DennunzioFGM2020TCS,GCA24,ManziniM99}.

For additive cellular automata over finite abelian groups, previous work provided a first structural characterization of strong transitivity, showing that it is preserved under spatial translations and temporal iterations
\cite{DennunzioFGM19,DennunzioFGM2020TCS}. That work explicitly raised the question of whether the same result extends to general cellular automata \cite{DennunzioFGM19}. 
Moreover, while the decidability of the weaker property of topological transitivity has been established for group cellular automata over arbitrary finite groups and in every dimension \cite{CDM2026TT}, a complete effective criterion for deciding strong transitivity from the local rule was still missing. Thus, determining strong transitivity remained a natural outstanding problem in the algorithmic study of group cellular automata.

In this paper we study strong transitivity for \emph{group cellular automata} (GCAs), \ie CAs whose alphabet is a finite group $G$ and whose global map is an endomorphism of the group $G^\zzd$. GCAs generalize linear CAs over $(\mathbb Z/m\mathbb Z)^n$, and their algebraic structure makes many dynamical properties accessible to exact analysis. Decidability results for GCAs have been obtained for injectivity and surjectivity and for several properties of the limit set~\cite{BeaurK24}, for topological transitivity~\cite{CDM2026TT}, for positive expansivity~\cite{DBLP:journals/access/DennunzioFM23}, and for topological entropy~\cite{CASTRONUOVO2026103749}. In these works, a central role is played by a decomposition of a GCA into simpler components via quotients by suitable subgroups of the alphabet, which reduces the analysis to alphabets that are direct products of isomorphic simple groups.
 
\medskip\noindent
\textbf{Our contributions.}
\begin{itemize}
\item We prove that, in every dimension, no GCA over a non-abelian group is strongly transitive (Theorem~\ref{thm:no_st_nonabelian}). Hence the alphabet of a strongly transitive GCA is abelian.
\item For GCAs over abelian groups we introduce a decomposition via quotients by power subgroups $G^m=\{g^m:\ g\in G\}$, and we prove that it preserves strong transitivity: a GCA is strongly transitive if and only if all its components are (Theorem~\ref{thm:st_decomposition} and Corollary~\ref{cor:power_decomp}). Such components are linear CAs over groups of the form $(\mathbb Z/p\mathbb Z)^n$. We point out that similar decompositions are used in \cite{CASTRONUOVO2026103749} and in \cite{CDM2026TT}.
 \item For a linear CA over $(\mathbb Z/p\mathbb Z)^n$ with matrix $A$ of Laurent polynomials in $d$ variables, we prove that strong transitivity holds if and only if $\det(A)\neq0$ and no monic irreducible factor of the characteristic polynomial $\chi_A$ has a unit constant term (Theorem~\ref{thm:abelian_st}). The proof combines Pontryagin duality, which translates strong transitivity into the triviality of the eventual image of the dual endomorphism (Corollary~\ref{cor:st_dual}), with an algebraic characterization of the eventual image of a matrix over a noetherian integrally closed domain (Theorem~\ref{thm:eventual_image}).
\item As a consequence, strong transitivity is decidable for GCAs in every dimension (Corollaries~\ref{cor:st_characterization} and~\ref{cor:st_decidable}). 
We give an explicit decision procedure which, for every fixed dimension $d$, runs in time polynomial in the group size $N=|G|$, the number $k$ of neighbours, and  the neighbourhood radius $\varrho$, including the cost of factoring at most $\log_2 N$ multivariate polynomials over finite fields (Theorem~\ref{thm:st_complexity}
and the subsequent complexity analysis). Unlike the procedure for deciding topological transitivity in~\cite{CDM2026TT}, it does not require a search over arbitrary verbal subgroups: it immediately rejects non-abelian alphabets and decomposes abelian ones by means of power subgroups.

\end{itemize}
 
\medskip\noindent
\textbf{Organization of the paper.}
Section~\ref{sec:prelim} recalls the basic definitions and a characterization of strong transitivity for GCAs. Section~\ref{sec:nonabelian} shows that GCAs over non-abelian groups are never strongly transitive. Section~\ref{sec:decomposition} presents the decomposition of GCAs over abelian groups via power subgroups and proves that it preserves strong transitivity, and Section~\ref{sec:abelian} deals with linear CAs over $(\mathbb Z/p\mathbb Z)^n$, where the characterization via Pontryagin duality is obtained. Section~\ref{sec:complexity} analyzes the computational complexity of deciding strong transitivity.
Section~\ref{sec:conclusions} contains some concluding remarks and open problems.

\section{Preliminaries}\label{sec:prelim}

\subsection{Cellular automata and group cellular automata}

Let $d\geq1$ and let $A$ be a finite set. A \emph{configuration} is an element of $A^\zzd$; for $c\in A^\zzd$ and $v\in\zzd$ we write $c_v$ for the value of $c$ at $v$, and for $S\subseteq\zzd$ we write $c|_S$ for the restriction of $c$ to $S$. We endow $A^\zzd$ with the prodiscrete topology, which makes it a compact metrizable space. For a finite set $S\subseteq\zzd$ and $p\in A^S$, the \emph{cylinder}
\[
\Cyl_A(S,p)=\{c\in A^\zzd:\ c|_S=p\}
\]
is clopen, and cylinders form a basis of the topology. For $\ell\geq0$ we let $B_\ell=[-\ell,\ell]^d\cap\zzd$. Every non-empty open set contains a cylinder of the form $\Cyl_A(B_\ell,p)$.
For $u\in\zzd$, the shift $\sigma^u:A^\zzd\to A^\zzd$ is defined by $\sigma^u(c)_v=c_{v+u}$. A ($d$-dimensional) \emph{CA} is a continuous map $\glorule:A^\zzd\to A^\zzd$ commuting with every shift. By the Curtis--Hedlund--Lyndon theorem, $\glorule$ is given by a \emph{local rule} $f:A^k\to A$ and a set $V_f=\{v_1,\dots,v_k\}\subseteq\zzd$ of pairwise distinct vectors, the \emph{neighborhood}, via $\glorule(c)_v=f(c_{v+v_1},\dots,c_{v+v_k})$.

Let $G$ be a finite group with identity $e$. Then $G^\zzd$ is a compact group under the componentwise operation, with identity $e^\zzd$. A CA $\glorule:G^\zzd\to G^\zzd$ is a \emph{GCA} if it is a group endomorphism of $G^\zzd$. In this case the local rule is a homomorphism $f:G^k\to G$, and $f(g_1,\dots,g_k)=h_1(g_1)\cdots h_k(g_k)$ for endomorphisms $h_1,\dots,h_k$ of $G$ with pairwise commuting images~\cite{GCA24}; we write $f=(h_1,\dots,h_k)$.

\subsection{Strong transitivity}

\begin{definition}\label{def:st}
Let $X$ be a compact metric space. A continuous map $f:X\to X$ is \emph{strongly transitive} if $\bigcup_{n\geq0}f^n(U)=X$ for every non-empty open set $U\subseteq X$; equivalently, if $\bigcup_{n\geq0}f^{-n}(x)$ is dense in $X$ for every $x\in X$.
\end{definition}

If $X$ has at least two points, a strongly transitive map $f$ is surjective: if $x\notin f(X)$, then $x\notin\bigcup_{n\geq0}f^n(U)$ for every non-empty open set $U$ not containing $x$.

\begin{theorem}\label{thm:semiconj}
Let $X$ and $Y$ be compact metric spaces, and let $f:X\to X$ and $g:Y\to Y$ be continuous. Assume that there is a continuous surjection $h:X\to Y$ with $h\circ f=g\circ h$. If $f$ is strongly transitive, then $g$ is strongly transitive.
\end{theorem}
\begin{proof}
Let $U\subseteq Y$ be non-empty and open, and let $\bar y\in Y$. Choose $\bar x\in X$ with $h(\bar x)=\bar y$. Since $h^{-1}(U)$ is non-empty and open and $f$ is strongly transitive, there are $n\geq0$ and $x\in h^{-1}(U)$ with $f^n(x)=\bar x$. Then $\bar y=h(f^n(x))=g^n(h(x))\in g^n(U)$.
\end{proof}

The following lemma generalizes~\cite[Lemma~15]{DennunzioFGM2020TCS}.


\begin{lemma}\label{lem:st_equiv}
Let $\glorule$ be a $d$-dimensional GCA over a finite group $G$. Then $\glorule$ is strongly transitive if and only if:

\smallskip
\noindent $(1)\ $ $\glorule$ is surjective, and

\noindent $(2)\ $ for every $\ell\geq0$ and every $p\in G^{B_\ell}$ there exists $n\geq0$ such that $e^\zzd\in\glorule^n(\Cyl_G(B_\ell,p))$.

\smallskip
\noindent
Moreover, if $\glorule$ is strongly transitive, then for every $\ell\geq0$ there is $n_\ell\geq0$ such that $\glorule^{n_\ell}(\Cyl_G(B_\ell,p))=G^\zzd$ for every $p\in G^{B_\ell}$.
\end{lemma}

\begin{proof}
The ``only if'' part follows from the definition and from the remark after Definition~\ref{def:st}. Assume that (1) and (2) hold, and fix $\ell\geq0$. For $p\in G^{B_\ell}$ let $n_p$ be the least $n$ such that $e^\zzd\in\glorule^{n}(\Cyl_G(B_\ell,p))$. Since $\glorule(e^\zzd)=e^\zzd$, we have $e^\zzd\in\glorule^{n}(\Cyl_G(B_\ell,p))$ for every $n\geq n_p$. Let $n_\ell=\max\{n_p:\ p\in G^{B_\ell}\}$; then
\begin{equation}\label{eq_str_tr}
e^\zzd\in\glorule^{n_\ell}(\Cyl_G(B_\ell,p))\qquad\text{for every }p\in G^{B_\ell}.
\end{equation}
Let $y\in G^\zzd$ and $p\in G^{B_\ell}$. Since $\glorule$ is surjective, there is $c\in G^\zzd$ with $\glorule^{n_\ell}(c)=y$. The set $c^{-1}\Cyl_G(B_\ell,p)$ is the cylinder $\Cyl_G(B_\ell,q)$ with $q_v=c_v^{-1}p_v$; by~\eqref{eq_str_tr} there is $c'\in\Cyl_G(B_\ell,q)$ with $\glorule^{n_\ell}(c')=e^\zzd$. Then $cc'\in\Cyl_G(B_\ell,p)$ and $\glorule^{n_\ell}(cc')=\glorule^{n_\ell}(c)\glorule^{n_\ell}(c')=y$. This proves the ``moreover'' part, and, since every non-empty open set contains a cylinder $\Cyl_G(B_\ell,p)$, that $\glorule$ is strongly transitive.
\end{proof}

\section{Non-abelian components}\label{sec:nonabelian}
In this section we prove that there are no strongly transitive d-GCA over a non-abelian group. This extends the same result for $d=1$ proved in \cite{CASTRONUOVO2026103749}.

For a group $\Gamma$ we denote by $Z_\Gamma$ its center; note that $Z_{G^\zzd}=Z_G^\zzd$.

\begin{lemma}\label{lem:KerZ}
Let $\glorule$ be a surjective $d$-GCA over a finite group $G$. Then $\Ker(\glorule)\subseteq Z_G^\zzd$.
\end{lemma}
\begin{proof}
For $b,c\in G^\zzd$ let $[b,c]=b^{-1}c^{-1}bc$. Let $c\in\Ker(\glorule)$ and suppose that $c_v\notin Z_G$ for some $v\in\zzd$. Choose $g\in G$ with $[c_v,g]\neq e$, and let $b\in G^\zzd$ be the configuration with $b_v=g$ and $b_w=e$ for $w\neq v$. Since $\Ker(\glorule)$ is normal, $[c,b]\in\Ker(\glorule)$; moreover, $[c,b]$ is trivial outside $v$ and $[c,b]_v=[c_v,g]\neq e$. Thus $[c,b]$ and $e^\zzd$ are distinct finite configurations with the same image, so $\glorule$ is not pre-injective, and hence not surjective by the Garden of Eden theorem~\cite{CeccheriniSilbersteinCoornart2010}, a contradiction.
\end{proof}

\begin{theorem}\label{thm:no_st_nonabelian}
Let $\glorule$ be a $d$-GCA over a finite non-abelian group $G$. Then $\glorule$ is neither strongly transitive nor positively expansive.
\end{theorem}
\begin{proof}
Assume that $\glorule$ is strongly transitive. Then $\glorule$ is surjective, and so is $\glorule^m$ for every $m\geq1$; by Lemma~\ref{lem:KerZ}, $\Ker(\glorule^m)\subseteq Z_G^\zzd$ for every $m$. Since $G$ is non-abelian, there is $g\in G\setminus Z_G$; let $U=\{c\in G^\zzd:\ c_0=g\}$, a non-empty open set disjoint from $Z_G^\zzd$. Then $e^\zzd\notin\glorule^m(U)$ for every $m\geq0$, since otherwise some element of $U$ would lie in $\Ker(\glorule^m)\subseteq Z_G^\zzd$. This contradicts strong transitivity. For the second assertion, if $d=1$ it suffices to recall that every positively expansive CA is strongly transitive~\cite[Theorem~2]{BlanchardCervelleFormenti2005}, while for $d\geq2$ there are no positively expansive CAs at all~\cite[Section~8]{Kari2005survey}.
\end{proof}

\section{A decomposition that preserves strong transitivity}\label{sec:decomposition}

By Theorem~\ref{thm:no_st_nonabelian}, a GCA over a non-abelian group is never strongly transitive. Hence, in order to decide strong transitivity, we may always assume that the alphabet is abelian, and in this section we develop a decomposition of GCAs over abelian groups which preserves strong transitivity.

\subsection{Quotients by invariant subgroups}

Let $G$ be a finite group and let $\glorule$ be a $d$-GCA on $G$ with local rule $f=(h_1,\dots,h_k)$. Let $H$ be a normal subgroup of $G$ such that $\glorule(H^\zzd)\subseteq H^\zzd$; this holds, for instance, if $h_j(H)\subseteq H$ for every $j$. For $g\in G$ let $[g]=gH\in G/H$, and for $c\in G^\zzd$ let $[c]\in(G/H)^\zzd$ be defined by $[c]_v=[c_v]$. The maps
\[
\overline\glorule:H^\zzd\to H^\zzd,\ \ \overline\glorule(c)=\glorule(c),
\qquad
\widetilde\glorule:(G/H)^\zzd\to(G/H)^\zzd,\ \ \widetilde\glorule([c])=[\glorule(c)],
\]
are well defined GCAs on $H$ and on $G/H$, respectively: $\widetilde\glorule$ does not depend on the representative because $\glorule(cz)=\glorule(c)\glorule(z)$ with $\glorule(z)\in H^\zzd$ for every $z\in H^\zzd$.

\begin{proposition}\label{prop:st_up}
Let $H$ be a normal subgroup of $G$ with $\glorule(H^\zzd)\subseteq H^\zzd$. If $\widetilde\glorule$ and $\overline\glorule$ are strongly transitive, then $\glorule$ is strongly transitive.
\end{proposition}
\begin{proof}
\emph{$\glorule$ is surjective.}
(Notice that this same fact, when $d=1,$ has been proved in \cite{CASTRONUOVO2026103749}).
Let $y\in G^\zzd$. Since $\widetilde\glorule$ is surjective, there is $x\in G^\zzd$ with $[\glorule(x)]=[y]$, \ie $z:=\glorule(x)^{-1}y\in H^\zzd$. Since $\overline\glorule$ is surjective, $z=\glorule(z')$ for some $z'\in H^\zzd$, and $y=\glorule(xz')$.

\emph{Condition (2) of Lemma~\ref{lem:st_equiv}.} Let $\ell\geq0$, $p\in G^{B_\ell}$, and $B=\Cyl_G(B_\ell,p)$. Let $\widetilde B=\Cyl_{G/H}(B_\ell,\widetilde p)$, where $\widetilde p_v=[p_v]$. Since $\widetilde\glorule$ is strongly transitive, there is $\widetilde n$ with $[e]^\zzd\in\widetilde\glorule^{\widetilde n}(\widetilde B)$, that is, there are $b\in B$ and $h\in H^\zzd$ with $\glorule^{\widetilde n}(b)=h$. The map $\glorule^{\widetilde n}$ is a continuous surjective endomorphism of the compact group $G^\zzd$, hence it is open by the open mapping theorem for compact groups~\cite[Theorem~5.29]{HewittRoss1963}. Thus $\glorule^{\widetilde n}(B)$ is an open set containing $h$, and it contains a cylinder $C$ with $h\in C$. The set $C\cap H^\zzd$ is non-empty and open in $H^\zzd$; since $\overline\glorule$ is strongly transitive, there are $\overline n$ and $a\in C\cap H^\zzd$ with $\glorule^{\overline n}(a)=e^\zzd$. Writing $a=\glorule^{\widetilde n}(b')$ with $b'\in B$, we get $\glorule^{\widetilde n+\overline n}(b')=e^\zzd$. By Lemma~\ref{lem:st_equiv}, $\glorule$ is strongly transitive.
\end{proof}

\subsection{Power subgroups of abelian groups}

From now on in this section $G$ is a finite abelian group.
For a prime $p$, a finite group is a \emph{$p$-group} if its order is a power of $p$, and a finite abelian group is an \emph{elementary abelian $p$-group} if it is isomorphic to $(\ZZ/p\ZZ)^n$ for some $n\geq0$, equivalently if $g^p=e$ for every element $g$.
For an integer $m\geq0$, the \emph{power subgroup} $G^m=\{g^m:\ g\in G\}$ is a subgroup of $G$, since $g^mh^m=(gh)^m$. Every endomorphism $\varphi$ of $G$ satisfies $\varphi(G^m)=\varphi(G)^m\subseteq G^m$; in particular $h_j(G^m)\subseteq G^m$ for every $j$, so that $\glorule((G^m)^\zzd)\subseteq(G^m)^\zzd$, and the GCAs $\overline\glorule$ on $G^m$ and $\widetilde\glorule$ on $G/G^m$ are defined.



\begin{theorem}\label{thm:st_decomposition}
Let $\glorule$ be a $d$-GCA over a finite abelian group $G$, and let $m\geq0$. Then $\glorule$ is strongly transitive if and only if both $\overline\glorule$ on $G^m$ and $\widetilde\glorule$ on $G/G^m$ are strongly transitive.
\end{theorem}
\begin{proof}
Let $H=G^m$. The ``if'' part is Proposition~\ref{prop:st_up}. Assume that $\glorule$ is strongly transitive.

\emph{$\widetilde\glorule$ is strongly transitive.} The map $\pi:G^\zzd\to(G/H)^\zzd$, $c\mapsto[c]$, is a continuous surjection with $\pi\circ\glorule=\widetilde\glorule\circ\pi$; apply Theorem~\ref{thm:semiconj}.

\emph{$\overline\glorule$ is strongly transitive.} Let $\ell\geq0$, $p\in H^{B_\ell}$ and $y\in H^\zzd$. Since the cylinders $\Cyl_H(B_\ell,p)=\Cyl_G(B_\ell,p)\cap H^\zzd$ form a basis of the topology of $H^\zzd$, it suffices to find $n\geq0$ and $c\in\Cyl_H(B_\ell,p)$ with $\glorule^n(c)=y$. Since $H=G^m$, there are $a\in G^{B_\ell}$ and $b\in G^\zzd$ with $p_u=a_u^m$ for $u\in B_\ell$ and $y_u=b_u^m$ for $u\in\zzd$. Since $\glorule$ is strongly transitive, there are $n\geq0$ and $c'\in\Cyl_G(B_\ell,a)$ with $\glorule^n(c')=b$. Let $c=c'^m$, the power being taken in the group $G^\zzd$, \ie cell by cell. Then $c\in H^\zzd$, $c|_{B_\ell}=p$, and, since $\glorule^n$ is an endomorphism, $\glorule^n(c)=\glorule^n(c')^m=b^m=y$.
\end{proof}

\subsection{The decomposition}


\begin{algorithm}[ht]
\caption{Abelian GCA decomposition}\label{alg:powerdecomp}
\KwIn{An abelian GCA $(\glorule, G)$}
 \KwOut{A finite set $\{(\glorule_1,G_1),\dots,(\glorule_q,G_q)\}$ of GCA where each $G_i$ is an elementary abelian $p_i$-group, or $G_i=G=\{e\}$}
\SetKwFunction{Decomposition}{PowerDecomp}
\SetKwProg{Fn}{Function}{:}{}
\Fn{\Decomposition{$\glorule$,$G$}}
{
    \eIf{$G$ is non-trivial and is not an elementary abelian $p$-group}{
        $p\gets$ a prime divisor of $|G|$\;
        \KwRet{\Decomposition{$\overline{\glorule}, G^p$} $\cup$ \Decomposition{$\widetilde{\glorule}, G/G^p$}}
    }{
        \KwRet{$\{(\glorule, G)\}$}
    }
}
\end{algorithm}


Whenever a recursive call is made, $G$ is non-trivial and not an elementary abelian
$p$-group, and $p$ is a prime divisor of $|G|$; let $H=G^p$. By Cauchy's theorem, the
endomorphism $g\mapsto g^p$ of $G$ has a non-trivial kernel, so $H\neq G$; and
$H\neq\{e\}$, since otherwise $g^p=e$ for every $g\in G$, \ie $G$ would be an elementary
abelian $p$-group. Hence $|H|<|G|$ and $|G/H|<|G|$, and {\tt PowerDecomp} terminates.
Since subgroups and quotients of abelian groups are abelian, it returns a finite family
$\{(\glorule_1,G_1),\dots,(\glorule_q,G_q)\}$ of $d$-GCAs over abelian groups. If
$G=\{e\}$, the output is $\{(\glorule,G)\}$; otherwise, by the stopping condition, every
$G_i$ satisfies $G_i\cong(\ZZ/p_i\ZZ)^{n_i}$ for some prime $p_i$ and some $n_i\geq1$.

\begin{corollary}\label{cor:power_decomp}
Let $\glorule$ be a $d$-GCA over a finite abelian group $G\neq\{e\}$, and let\\
$\{(\glorule_1,G_1),\dots,(\glorule_q,G_q)\}$ be the output of {\tt PowerDecomp}
(Algorithm~\ref{alg:powerdecomp}) on $(\glorule,G)$. Then, for every $i$,
$G_i\cong(\ZZ/p_i\ZZ)^{n_i}$ for some prime $p_i$ and some $n_i\geq1$, and $\glorule$ is
strongly transitive if and only if every $\glorule_i$ is strongly transitive.
\end{corollary}
\begin{proof}
The first assertion follows from the stopping condition of {\tt PowerDecomp}, as observed
above; the second from Theorem~\ref{thm:st_decomposition} and induction on the depth of the recursion.
\end{proof}

Hence, by Theorem~\ref{thm:no_st_nonabelian} and Corollary~\ref{cor:power_decomp}, in order
to decide strong transitivity of GCAs it suffices to decide it for GCAs over groups of the
form $(\ZZ/p\ZZ)^n$, which are studied in Section~\ref{sec:abelian}.

\section{Linear GCAs over $(\mathbb Z/p\mathbb Z)^n$ }\label{sec:abelian}

Throughout this section the dimension $d\geq1$ is arbitrary. By Corollary~\ref{cor:power_decomp}, it remains to characterize strong transitivity for
GCAs over $(\ZZ/p\ZZ)^n$; as we show below, these are exactly the linear CAs over $\Fp^n$.

\subsection{Linear CAs and their matrices}

Let $p$ be a prime, $\Fp=\ZZ/p\ZZ$, $\LP=\Fp[x_1^{\pm1},\dots,x_d^{\pm1}]$ the ring of Laurent polynomials in $d$ variables, and $\KP=\Fp(x_1,\dots,x_d)$ its field of fractions. For $u\in\zzd$ we write $x^u=x_1^{u_1}\cdots x_d^{u_d}$. The ring $\LP$ is a noetherian unique factorization domain, hence integrally closed, and its units are the elements $ax^u$ with $a\in\Fp\setminus\{0\}$ and $u\in\zzd$. Every endomorphism of $\Fp^n$ is $\Fp$-linear, so a GCA $\glorule$ on $\Fp^n$ has the form $\glorule(c)_v=\sum_{u\in V_f}A_uc_{v+u}$ with $A_u\in M_n(\Fp)$. We associate with $\glorule$ the matrix $M_\glorule=\sum_{u\in V_f}A_ux^{-u}\in M_n(\LP)$; conversely, every $A\in M_n(\LP)$ defines in this way a GCA $\glorule_A$ on $\Fp^n$. Identifying a finite configuration $c$ with $\sum_vc_vx^v\in\LP^n$, $\glorule_A$ acts on finite configurations as the multiplication by $A$, and $\glorule_{AB}=\glorule_A\circ\glorule_B$. We say that $A$ is strongly transitive (resp.\ transitive) if $\glorule_A$ is.

\subsection{Strong transitivity via Pontryagin duality}

For Pontryagin duality we refer to~\cite{Morris1977}.

Let $X$ be a compact metrizable abelian group, written additively with identity $0$, and let $F:X\to X$ be a continuous endomorphism. Let
\[
\mathcal K(F)=\bigcup_{n\geq0}\Ker(F^n).
\]

\begin{lemma}\label{lem:st_group}
$F$ is strongly transitive if and only if $F$ is surjective and $\overline{\mathcal K(F)}=X$.
\end{lemma}
\begin{proof}
Assume that $F$ is strongly transitive. Then $F$ is surjective (see the remark after Definition~\ref{def:st}; if $X=\{0\}$ there is nothing to prove). For every non-empty open set $U$ there is $n$ with $0\in F^n(U)$, \ie $U\cap\Ker(F^n)\neq\emptyset$; hence $\mathcal K(F)$ is dense.

Conversely, assume that $F$ is surjective and $\mathcal K(F)$ is dense. Let $U\subseteq X$ be non-empty and open, let $u_0\in U$, and let $V$ be an open neighborhood of $0$ with $V=-V$ and $u_0+V\subseteq U$. For every $z\in X$ the open set $z-V$ meets $\mathcal K(F)$, so $X=\mathcal K(F)+V=\bigcup_{n\geq0}(\Ker(F^n)+V)$. The sets $\Ker(F^n)+V$ are open and increasing in $n$, so by compactness $X=\Ker(F^N)+V$ for some $N$. Let $x\in X$. Since $F^N$ is surjective, $x=F^N(y)$ for some $y\in X$; write $y-u_0=k+v$ with $k\in\Ker(F^N)$ and $v\in V$. Then $u_0+v\in U$ and $F^N(u_0+v)=F^N(y-k)=x$. Hence $x\in F^N(U)$.
\end{proof}

Let $\widehat X$ be the Pontryagin dual of $X$, the discrete abelian group of continuous homomorphisms $\chi:X\to\mathbb T$, where $\mathbb T$ is the circle group, written multiplicatively; $\widehat X$ is written multiplicatively, with identity the trivial character $1$. Let $\widehat F:\widehat X\to\widehat X$, $\widehat F(\chi)=\chi\circ F$, be the dual endomorphism. For $S\subseteq X$, the \emph{annihilator} of $S$ is the subgroup
\[
S^\perp=\{\chi\in\widehat X:\ \chi(s)=1\text{ for every }s\in S\}.
\]

\begin{lemma}\label{lem:ann_union}
For every family $(H_i)_{i\in I}$ of subsets of $X$, $\big(\bigcup_iH_i\big)^\perp=\bigcap_iH_i^\perp$.
\end{lemma}
\begin{proof}
A character annihilates $\bigcup_iH_i$ if and only if it annihilates each $H_i$.
\end{proof}

\begin{lemma}\label{lem:ann_ker}
For every continuous endomorphism $\varphi:X\to X$, $(\Ker\varphi)^\perp=\Imma(\widehat\varphi)$.
\end{lemma}
\begin{proof}
If $\psi\in\widehat X$, then $\widehat\varphi(\psi)=\psi\circ\varphi$ is trivial on $\Ker\varphi$; hence $\Imma(\widehat\varphi)\subseteq(\Ker\varphi)^\perp$. Conversely, let $\chi\in(\Ker\varphi)^\perp$. The subgroup $Y=\varphi(X)$ is compact, hence closed in $X$. Since $\chi$ is trivial on $\Ker\varphi$, the map $\psi_0:Y\to\mathbb T$, $\psi_0(\varphi(x))=\chi(x)$, is a well-defined homomorphism. It is continuous: for every closed set $C\subseteq\mathbb T$, $\psi_0^{-1}(C)=\varphi(\chi^{-1}(C))$ is the image of a compact set, hence closed. By~\cite[Corollary of Theorem~27]{Morris1977}, $\psi_0$ extends to a character $\psi$ of $X$, and $\widehat\varphi(\psi)=\psi\circ\varphi=\chi$.
\end{proof}

Applying Lemma~\ref{lem:ann_ker} to $\varphi=F^n$ and using $\widehat{F^n}=\widehat F^{\,n}$, Lemma~\ref{lem:ann_union} gives
\begin{equation}\label{eq:Kperp}
\mathcal K(F)^\perp=\bigcap_{n\geq0}\Ker(F^n)^\perp=\bigcap_{n\geq0}\Imma(\widehat F^{\,n}).
\end{equation}

\begin{lemma}\label{lem:dense_ann}
For every subgroup $H\leq X$, $\overline H=X$ if and only if $H^\perp=\{1\}$.
\end{lemma}
\begin{proof}
If $\overline H=X$ and $\chi\in H^\perp$, then $\chi$ is identically $1$ on $X$ by continuity. Conversely, let $A=\overline H^\perp$. By~\cite[Proposition~38]{Morris1977}, $\overline H$ is the annihilator of $A$ in $X$, \ie $\overline H=\{x\in X:\ \chi(x)=1\text{ for every }\chi\in A\}$. If $\overline H\neq X$, then $A\neq\{1\}$; since $H^\perp=\overline H^\perp$ by continuity, $H^\perp\neq\{1\}$.
\end{proof}

\begin{theorem}\label{thm:dual}
The following conditions are equivalent:
\begin{itemize}
\item[(1)] $\overline{\mathcal K(F)}=X$;
\item[(2)] $\bigcap_{n\geq0}\Imma(\widehat F^{\,n})=\{1\}$.
\end{itemize}
\end{theorem}
\begin{proof}
By Equation~\eqref{eq:Kperp} and Lemma~\ref{lem:dense_ann}.
\end{proof}

Let $\Lambda$ be an abelian group and $\psi:\Lambda\to\Lambda$ an endomorphism. The set of \emph{infinitely $\psi$-divisible} elements is
\[
\mathrm{Div}_\infty(\psi)=\{m\in\Lambda:\ \text{for every }n\geq0\text{ there is }x_n\in\Lambda\text{ with }m=\psi^n(x_n)\}=\bigcap_{n\geq0}\Imma(\psi^n).
\]
By Theorem~\ref{thm:dual} applied with $\psi=\widehat F$,
\[
\overline{\mathcal K(F)}=X\quad\Longleftrightarrow\quad\mathrm{Div}_\infty(\widehat F)=\{1\},
\]
that is, $\mathcal K(F)$ is dense if and only if there are no non-trivial infinitely $\widehat F$-divisible elements.

\begin{corollary}\label{cor:st_dual}
$F$ is strongly transitive if and only if $\widehat F$ is injective and $\mathrm{Div}_\infty(\widehat F)=\{1\}$.
\end{corollary}
\begin{proof}
A character $\chi$ satisfies $\widehat F(\chi)=\chi\circ F=1$ if and only if $\chi$ annihilates $F(X)$; thus $\Ker(\widehat F)=F(X)^\perp$. Since $F(X)$ is compact, hence closed, Lemma~\ref{lem:dense_ann} gives that $F$ is surjective if and only if $\widehat F$ is injective. The claim follows from Lemma~\ref{lem:st_group} and Theorem~\ref{thm:dual}.
\end{proof}

\subsection{The eventual image of a matrix}

Throughout this subsection $L$ is a noetherian integrally closed domain and $K=\mathrm{Frac}(L)$; for $M\in M_n(L)$ we let $\chi_M(t)=\det(tI-M)$, and $M$ acts on $L^n$ (resp.\ $K^n$) by left multiplication. We shall use the following facts.
\begin{itemize}
\item[(F1)] If $f\in L[t]$ is monic and $f=gh$ with $g,h\in K[t]$ monic, then $g,h\in L[t]$~\cite[Chapter~5, Exercise~8(i)]{atiyah1969introduction}. In particular $g(0)h(0)=f(0)$ in $L$, so that $g(0)$ is a unit of $L$ whenever $f(0)$ is.
\item[(F2)] The multiplication by $M\in M_n(L)$ is injective on $L^n$ if and only if $\det(M)\neq0$. Indeed, if $\det(M)\neq0$, then $\mathrm{adj}(M)M=\det(M)I$ shows injectivity, $L$ being a domain; if $\det(M)=0$, there is $0\neq v\in K^n$ with $Mv=0$, and multiplying $v$ by a common denominator of its entries gives a non-zero element of the kernel in $L^n$.
\end{itemize}

\begin{theorem}\label{thm:eventual_image}
Let $M\in M_n(L)$ with $\det(M)\neq0$, and let
\[
N=\bigcap_{k\geq0}\Imma(M^k)\subseteq L^n.
\]
Then $N\neq\{0\}$ if and only if $\chi_M$ has a monic irreducible factor $g\in K[t]$ such that $g(0)$ is a unit of $L$.
\end{theorem}

\begin{proof}
Recall that every monic factor of $\chi_M$ in $K[t]$ belongs to $L[t]$ by (F1).

($\Rightarrow$) Assume that $N\neq\{0\}$.

\emph{$M$ restricts to an automorphism of $N$.} By (F2), $M$ is injective on $L^n$. If $x\in N$, then $Mx\in\Imma(M^{k+1})\subseteq\Imma(M^k)$ for every $k$, so $Mx\in N$. Conversely, let $x\in N$; for every $k\geq0$ write $x=M^{k+1}z_k$ with $z_k\in L^n$, and put $y_k=M^kz_k$. Then $My_k=x$ for every $k$, so by injectivity $y:=y_0=y_k\in\Imma(M^k)$ for every $k$, \ie $y\in N$ and $My=x$. Hence $M':N\to N$, $M'(x)=Mx$, is a bijective $L$-linear map.

\emph{The subspace $W$.} Regard $L^n$ as a subset of $K^n$, and let $W\subseteq K^n$ be the $K$-span of $N$. Since $N\neq\{0\}$, $r=\dim_KW\geq1$. Since $MN\subseteq N$, we have $MW\subseteq W$; let $\varphi:W\to W$ be the restriction of the multiplication by $M$. Since $\varphi(W)$ is spanned by $M'(N)=N$, $\varphi$ is surjective, hence invertible, and $\varphi^{-1}$ extends $M'^{-1}$.

\emph{$\chi_\varphi\in L[t]$ and $\det(\varphi)$ is a unit of $L$.} Since $L$ is noetherian, $N\subseteq L^n$ is finitely generated. The maps $M'$ and $M'^{-1}$ are $L$-linear endomorphisms of $N$, so by~\cite[Proposition~2.4]{atiyah1969introduction} each of them is annihilated by a monic polynomial with coefficients in $L$. Since $W$ is spanned by $N$, the same polynomials annihilate $\varphi$ and $\varphi^{-1}$, respectively. Hence every eigenvalue $\lambda$ of $\varphi$ in an algebraic closure $\overline K$ of $K$ is integral over $L$, and so is $\lambda^{-1}$, which is an eigenvalue of $\varphi^{-1}$. The coefficients of $\chi_\varphi$, as well as $\det(\varphi)$ and $\det(\varphi)^{-1}$, are polynomial expressions in such eigenvalues, hence elements of $K$ integral over $L$~\cite[Corollary~5.3]{atiyah1969introduction}. Since $L$ is integrally closed, $\chi_\varphi\in L[t]$ and $\det(\varphi)$ is a unit of $L$.

\emph{Conclusion.} Completing a basis of $W$ to a basis of $K^n$, the multiplication by $M$ is represented by a block upper triangular matrix whose upper left block represents $\varphi$; hence $\chi_\varphi$ divides $\chi_M$ in $K[t]$. Let $g$ be a monic irreducible factor of $\chi_\varphi$ in $K[t]$; then $g$ divides $\chi_M$. Since $\chi_\varphi(0)=(-1)^r\det(\varphi)$ is a unit of $L$, $g(0)$ is a unit of $L$ by (F1), applied to $f=\chi_\varphi$.

($\Leftarrow$) Let $g$ be a monic irreducible factor of $\chi_M$ with $g(0)$ a unit of $L$, and let $e\geq1$ be its multiplicity in $\chi_M$. Since $g$ divides $\chi_M$, $g$ has a root which is an eigenvalue of $M$, so that $g(M)$, and hence $g(M)^e$, is not injective on $K^n$; thus $W_g=\Ker\big(g(M)^e\big)\subseteq K^n$ is a non-zero $K$-subspace, and $MW_g\subseteq W_g$. By (F1), $g\in L[t]$, so we can write $g(t)^e=t\,q(t)+c$ with $q\in L[t]$ and $c=g(0)^e$ a unit of $L$. On $W_g$ we have $Mq(M)=-c\,\mathrm{id}$, that is, the restriction of $M$ to $W_g$ is invertible with inverse the restriction of $B=-c^{-1}q(M)\in M_n(L)$. Let $v\in W_g\cap L^n$ be non-zero; it exists, since a non-zero element of $W_g$ multiplied by a common denominator of its entries lies in $W_g\cap L^n$. For every $k\geq0$,
since $B$ is a polynomial in $M$ and $MW_g\subseteq W_g,$ $B^kv\in L^n\cap W_g$ and $M^k(B^kv)=v$. Hence $v\in\Imma(M^k)$ for every $k$, \ie $0\neq v\in N$.
\end{proof}

\subsection{Characterization of strongly transitive linear GCAs}

Let $\sigma:\LP\to\LP$ be the ring automorphism defined by $\sigma(x_i)=x_i^{-1}$ for $1\leq i\leq d$; it extends to an automorphism of $\KP$, and, acting on coefficients, to automorphisms of $M_n(\LP)$ and of $\KP[t]$ fixing $t$. For $A\in M_n(\LP)$ let $A^*=\sigma(A)^T$; note that $(AB)^*=B^*A^*$, so $(A^k)^*=(A^*)^k$.

Let $\omega=e^{2\pi i/p}$. For $a=\sum_va_vx^v\in\LP^n$, with $a_v\in\Fp^n$ and finitely many $a_v\neq0$, let $\chi_a:(\Fp^n)^\zzd\to\mathbb T$ be defined by $\chi_a(c)=\omega^{\sum_va_v\cdot c_v}$, where $a_v\cdot c_v$ is the standard scalar product of $\Fp^n$.

\begin{lemma}\label{lem:dual_linear}
The map $a\mapsto\chi_a$ is an isomorphism from the additive group $\LP^n$ onto $\widehat{(\Fp^n)^\zzd}$. For every $A\in M_n(\LP)$ we have $\widehat{\glorule_A}(\chi_a)=\chi_{A^*a}$.
\end{lemma}
\begin{proof}
Clearly $\chi_a$ is a continuous character and $\chi_{a+b}=\chi_a\chi_b$. Let $\chi$ be a continuous character. Since $(\Fp^n)^\zzd$ has exponent $p$, $\chi$ takes values in the finite, hence discrete, subgroup $\{\omega^s:\ s\in\Fp\}$ of $\mathbb T$, so its kernel is open and contains all the configurations vanishing on some finite set $S\subseteq\zzd$. Hence $\chi(c)$ depends only on $c|_S$, and since every character of $\Fp^n$ has the form $g\mapsto\omega^{b\cdot g}$ for a unique $b\in\Fp^n$, we get $\chi=\chi_a$ for a unique $a\in\LP^n$. For the second claim, write $A=\sum_uA_ux^{-u}$, so that $\glorule_A(c)_v=\sum_uA_uc_{v+u}$. Then
\[
\sum_va_v\cdot\glorule_A(c)_v=\sum_w\Big(\sum_uA_u^Ta_{w-u}\Big)\cdot c_w,
\]
and $\sum_w\big(\sum_uA_u^Ta_{w-u}\big)x^w=\big(\sum_uA_u^Tx^u\big)a=A^*a$.
\end{proof}

Here we reprove the well known (see \cite{bruyn1991}) characterization of surjectivity for linear cellular automata. 

\begin{corollary}\label{cor:surj_det}
Let $A\in M_n(\LP)$. Then $\glorule_A$ is surjective if and only if $\det(A)\neq0$.
\end{corollary}
\begin{proof}
By the proof of Corollary~\ref{cor:st_dual}, $\glorule_A$ is surjective if and only if $\widehat{\glorule_A}$ is injective, that is, by Lemma~\ref{lem:dual_linear}, if and only if the multiplication by $A^*$ is injective on $\LP^n$. By (F2), this holds if and only if $\det(A^*)=\sigma(\det(A))\neq0$, \ie $\det(A)\neq0$.
\end{proof}

\begin{theorem}\label{thm:abelian_st}
Let $A\in M_n(\LP)$, with characteristic polynomial $\chi_A(t)=\det(tI-A)$. Then $\glorule_A$ is strongly transitive if and only if $\det(A)\neq0$ and, for every monic irreducible factor $g$ of $\chi_A$ in $\KP[t]$, $g(0)$ is not a unit of $\LP$.
\end{theorem}
\begin{proof}
Recall that $\LP$ is a noetherian integrally closed domain, so that (F1), (F2) and Theorem~\ref{thm:eventual_image} apply with $L=\LP$. By Corollary~\ref{cor:st_dual} and Lemma~\ref{lem:dual_linear}, under which $\widehat{\glorule_A}^{\,k}=\widehat{\glorule_{A^k}}$ corresponds to the multiplication by $(A^*)^k$, $\glorule_A$ is strongly transitive if and only if the multiplication by $A^*$ is injective on $\LP^n$ and $\bigcap_{k\geq0}\Imma((A^*)^k)=\{0\}$. By (F2), the first condition is equivalent to $\det(A^*)=\sigma(\det(A))\neq0$, \ie to $\det(A)\neq0$. Assume $\det(A)\neq0$. By Theorem~\ref{thm:eventual_image}, the second condition holds if and only if no monic irreducible factor of $\chi_{A^*}$ has a unit constant term. Now $\chi_{A^*}(t)=\det(tI-\sigma(A)^T)=\sigma(\chi_A)(t)$, and $\sigma$ is a ring automorphism of $\KP[t]$ preserving $\LP$ and its units; hence the monic irreducible factors of $\chi_{A^*}$ are the polynomials $\sigma(g)$, with $g$ a monic irreducible factor of $\chi_A$, and $\sigma(g)(0)=\sigma(g(0))$ is a unit if and only if $g(0)$ is.
\end{proof}

\begin{corollary}\label{cor:st_characterization}
Let $\glorule$ be a $d$-GCA over a finite group $G$. Then $\glorule$ is strongly transitive if and only if $G$ is abelian and, denoting by $\{(\glorule_1,G_1),\dots,(\glorule_q,G_q)\}$ the output of {\tt PowerDecomp} on $(\glorule,G)$, for every $i$ with $G_i\neq\{e\}$ the following holds: $G_i\cong\mathbb F_{p_i}^{n_i}$, and the matrix $A_i\in M_{n_i}(\mathbb L_{p_i})$ of the GCA on $\mathbb F_{p_i}^{n_i}$ corresponding to $\glorule_i$ under such an isomorphism satisfies $\det(A_i)\neq0$, and $g(0)$ is not a unit of $\mathbb L_{p_i}$ for every monic irreducible factor $g$ of $\chi_{A_i}$ in $\mathbb K_{p_i}[t]$.
\end{corollary}
\begin{proof}
If $G$ is non-abelian, $\glorule$ is not strongly transitive by Theorem~\ref{thm:no_st_nonabelian}. Let $G$ be abelian. By Corollary~\ref{cor:power_decomp}, $\glorule$ is strongly transitive if and only if every $\glorule_i$ is. If $G_i=\{e\}$ (which happens only if $G=\{e\}$), then $G_i^\zzd$ is a singleton and $\glorule_i$ is strongly transitive. Otherwise $G_i\cong\mathbb F_{p_i}^{n_i}$ with $n_i\geq1$; an isomorphism $\psi:G_i\to\mathbb F_{p_i}^{n_i}$ induces, cell by cell, a homeomorphism $\Psi:G_i^\zzd\to(\mathbb F_{p_i}^{n_i})^\zzd$ conjugating $\glorule_i$ to the GCA $\glorule_{A_i}=\Psi\circ\glorule_i\circ\Psi^{-1}$. By Theorem~\ref{thm:semiconj}, applied to $\Psi$ and to $\Psi^{-1}$, $\glorule_i$ is strongly transitive if and only if $\glorule_{A_i}$ is, and the latter is characterized by Theorem~\ref{thm:abelian_st}.
\end{proof}

\begin{corollary}\label{cor:st_decidable}
For every fixed $d\geq1$, strong transitivity is decidable for $d$-GCAs over finite groups: there is an algorithm which, given a finite group $G$ by its multiplication table and the local rule of a $d$-GCA $\glorule$ on $G$, decides whether $\glorule$ is strongly transitive.
\end{corollary}
\begin{proof}
Whether $G$ is abelian is decided from its multiplication table; if it is not, the answer is negative by Corollary~\ref{cor:st_characterization}. Otherwise, the output of {\tt PowerDecomp} on $(\glorule,G)$ is computable, since power subgroups and quotients are computed from the multiplication table. For every $G_i\neq\{e\}$, an isomorphism $G_i\cong\mathbb F_{p_i}^{n_i}$ is found by exhaustive search, and the matrix $A_i$ is computed from the local rule of $\glorule_i$. The determinant and the characteristic polynomial of $A_i$ are computed in $\mathbb L_{p_i}$ and $\mathbb L_{p_i}[t]$. Multiplying $\chi_{A_i}$ by a suitable monomial $x^w$ yields a polynomial in $\mathbb F_{p_i}[x_1,\dots,x_d,t]$, whose factorization into irreducibles is effectively computable \cite{Lenstra1985,vonzurGathenGerhard2013}; by Gauss's lemma, this factorization yields the monic irreducible factors of $\chi_{A_i}$ in $\mathbb K_{p_i}[t]$. Finally, an element of $\mathbb L_{p_i}$ is a unit if and only if it is a single monomial $ax^u$ with $a\neq0$, which is checked directly. By Corollary~\ref{cor:st_characterization}, these computations decide whether $\glorule$ is strongly transitive. An explicit procedure, together with a bound on its cost, is given in Section~\ref{sec:complexity}.
\end{proof}

\section{Computational complexity}\label{sec:complexity}

In this section we give an explicit decision procedure for strong transitivity and bound its cost. By Theorem~\ref{thm:no_st_nonabelian}, a GCA over a non-abelian group is never strongly transitive; hence only GCAs over abelian groups need to be decomposed, and for them we use the procedure {\tt PowerDecomp}.

\medskip\noindent
\textbf{Input and cost model.}
The group $G$, of order $N$, is given by its multiplication table.
The $d$-GCA $F$ is given by the tables of the endomorphisms
$h_1,\ldots,h_k$ of its local rule and by its neighbourhood
$V_f=\{v_1,\ldots,v_k\}\subseteq\mathbb{Z}^d$.
Let $\varrho=\max_j\|v_j\|_\infty$, where
$\|u\|_\infty=\max_h |u_h|$ for $u\in\mathbb{Z}^d$.
We express the running-time bound in terms of the parameters
$N$, $k$, $d$, and $\varrho$.
We count as one operation a lookup in a table, an arithmetic operation
in $\mathbb{F}_p$, and an addition or comparison of integers.


\medskip\noindent
\textbf{The decision procedure.}
The decision procedure {\tt IsST} is given in Algorithm~\ref{alg:isst}: it rejects
non-abelian alphabets, decomposes abelian ones by {\tt PowerDecomp}
(Algorithm~\ref{alg:powerdecomp}), and applies to every component the function
{\tt AbelianTest} described below.

\begin{algorithm}[ht]
\caption{Testing whether a GCA is strongly transitive}\label{alg:isst}
\KwIn{A GCA $(\glorule, G)$}
 \KwOut{{\tt true} if $(\glorule, G)$ is strongly transitive, {\tt false} otherwise}
\SetKwFunction{IsST}{IsST}
\SetKwFunction{Decomposition}{PowerDecomp}
\SetKwFunction{AbTest}{AbelianTest}
\SetKwProg{Fn}{Function}{:}{}
\Fn{\IsST{$\glorule$, $G$}}{
    \If{$G$ is non-abelian}{\KwRet{\tt false}\;}
    \If{$G=\{e\}$}{\KwRet{\tt true}\;}

    $A=$\Decomposition{$\glorule$, $G$};\\
    \ForEach{$(\glorule_i,G_i)\in A$ }{
        \If{{\bf not} \AbTest{$\glorule_i$, $G_i$}}{\KwRet{\tt false}\;}
    }
    \KwRet{\tt true}\;
}
\end{algorithm}

The function {\tt AbelianTest}, applied to a GCA $\glorule$ over $G\cong\Fp^n$, computes an isomorphism $G\cong\Fp^n$, the matrix $A\in M_n(\LP)$ of the corresponding GCA on $\Fp^n$, and the polynomial $\chi_A$; it returns {\tt true} if and only if $\det(A)\neq0$ and no monic irreducible factor of $\chi_A$ in $\KP[t]$ has a unit constant term.

\begin{lemma}\label{lem:isst_correct}
{\tt IsST} terminates on every input $(\glorule,G)$, and it returns {\tt true} if and only if $\glorule$ is strongly transitive.
\end{lemma}
\begin{proof}
{\tt PowerDecomp} terminates (see the discussion after its definition), and so does {\tt IsST}.
If $G$ is non-abelian, $\glorule$ is not strongly transitive by Theorem~\ref{thm:no_st_nonabelian}.
If $G=\{e\}$, $G^\zzd$ is a singleton and $\glorule$ is strongly transitive.
Otherwise every $G_i$ in the output of {\tt PowerDecomp} satisfies $G_i\cong\mathbb F_{p_i}^{n_i}$ with $n_i\geq1$,
and the claim follows from Corollary~\ref{cor:power_decomp}, Theorem~\ref{thm:abelian_st}
and the conjugation argument in the proof of Corollary~\ref{cor:st_characterization}.
\end{proof}


\begin{lemma}\label{lem:st_coordinates}
Let $G\cong\Fp^n$ be given by its multiplication table, with $M=|G|=p^n$, and let $h_1,\dots,h_k$ be endomorphisms of $G$, given by their tables. One can compute $p$, a basis $g_1,\dots,g_n$ of $G$ over $\Fp$, a table of the coordinates of all the elements of $G$ in this basis, and the matrices in $M_n(\Fp)$ of $h_1,\dots,h_k$ in this basis, in $O(M\log M+kM)$ operations.
\end{lemma}
\begin{proof}
The prime $p$ is the order of any element $g\neq e$, computed in $O(p)\subseteq O(M)$ operations. Let $V_0=\{e\}$. Given the span $V_i$ of $g_1,\dots,g_i$, together with the coordinates of its elements, choose $g_{i+1}\notin V_i$ by scanning $G$ with a membership array. Then $V_{i+1}=\bigcup_{a\in\Fp}g_{i+1}^aV_i$, and the coordinates of its elements are obtained from those of $V_i$, in $O(n|V_{i+1}|+M)$ operations. Since $|V_{i+1}|=p^{i+1}$ and $n\leq\log_2M$, the whole construction costs $O(nM)=O(M\log M)$. The $j$-th column of the matrix of $h_l$ is the coordinate vector of $h_l(g_j)$, read from the tables in $O(n)$ operations. Reading the $k$ tables costs $O(kM)$, which dominates the $O(kn^2)$ operations needed to fill the matrices.
\end{proof}

\medskip\noindent
\textbf{The test of commutativity.}
Testing whether $G$ is abelian costs $O(N^2)$ operations, and it is performed only once.

\medskip\noindent
\textbf{Size of the recursion tree of {\tt PowerDecomp}.}
At every internal node the group $G$ is split into two non-trivial groups $H$ and $G/H$ with $|H|\,|G/H|=|G|$. Hence the orders of the groups attached to the leaves multiply to at most $N$, the tree has at most $\log_2N$ leaves and at most $2\log_2N$ nodes, and its depth is at most $\log_2N$. Moreover, the orders of the groups attached to the nodes at a fixed depth multiply to at most $N$.

\medskip\noindent
\textbf{Cost at an internal node.}
Let $M=|G|$. A prime divisor $p$ of $M$ is found in $O(\sqrt M)$ operations, and the powers $g^p$, $g\in G$, in $O(M\log p)$ operations by repeated squaring; they give both the subgroup $G^p$ and the test of whether $G$ is an elementary abelian $p$-group, \ie whether $M$ is a power of $p$ and $g^p=e$ for every $g\in G$. Building the multiplication tables of $H$ and $G/H$ costs $O(M^2)$, and the tables of the restrictions $h_j|_H$ and of the induced endomorphisms $[g]\mapsto[h_j(g)]$ of $G/H$ cost $O(kM)$; the neighborhood does not change. Since the orders at a fixed depth multiply to at most $N$, the sum of $M^2+kM$ over the nodes at a fixed depth is at most $N^2+kN$, and all the internal nodes together cost
\[
O\big((N^2+kN)\log N\big).
\]

\medskip\noindent
\textbf{Cost at a leaf.}
At a leaf, {\tt AbelianTest} is called on $G\cong\Fp^n$, with $M=|G|=p^n$ and $n\leq\log_2M$. By Lemma~\ref{lem:st_coordinates}, the matrices $A_1,\dots,A_k\in M_n(\Fp)$ of $h_1,\dots,h_k$ are computed in $O(M\log M+kM)$ operations, and $A=\sum_jA_jx^{-v_j}$ in $O(kn^2d)$ operations. The polynomial $\chi_A$ is computed by Berkowitz's division-free algorithm~\cite{Berkowitz1984}, which uses $O(n^4)$ operations in $\LP$. All the intermediate quantities are polynomials of degree at most $n$ in the entries of $A$, hence Laurent polynomials with exponents in $[-n\varrho,n\varrho]^d$; each of them has at most $(2n\varrho+1)^d$ monomials, and each operation in $\LP$ costs $O\big(d(2n\varrho+1)^{2d}\big)$. Then $\det(A)=(-1)^n\chi_A(0)$.

If $\det(A)\neq0$, let $w=(n\varrho,\dots,n\varrho)\in\zzd$ and $P=x^w\chi_A$. Then $P\in\Fp[x_1,\dots,x_d,t]$, with $\deg_tP=n$ and $\deg_{x_h}P\leq2n\varrho$ for every $h$. Let $T_{\rm fact}(p,n,n\varrho,d)$ be the cost of factoring such a polynomial into irreducibles. By Gauss's lemma, the irreducible factors of $P$ of positive degree in $t$ have monomial leading coefficients in $t$, since they divide the leading coefficient $x^w$ of $P$; dividing them by these coefficients yields the monic irreducible factors of $\chi_A$ in $\KP[t]$. Testing whether the constant term of each factor is a single monomial, \ie a unit of $\LP$, costs time linear in the size of the factorization.

\medskip\noindent
\textbf{Overall complexity.}
Let $M_1,\dots,M_q$ be the orders of the groups at the leaves of {\tt PowerDecomp}, and $n_i=\log_{p_i}M_i$. Since $\prod_iM_i\leq N$ and $M_i\geq2$, we have $\sum_iM_i\log M_i\leq N\log N$ and $\sum_in_i\leq\log_2N$; hence
\[
\sum_in_i^4(2n_i\varrho+1)^{2d}\leq(\log_2N)^4(2\varrho\log_2N+1)^{2d}.
\]
Adding the costs of the internal nodes and of the leaves, we obtain the following result.

\begin{theorem}\label{thm:st_complexity}
Let $\glorule$ be a $d$-GCA over a group of order $N$, with $k$ neighbors and $\varrho=\max_j\|v_j\|_\infty$. The procedure {\tt IsST} decides whether $\glorule$ is strongly transitive using
\[
O\Big((N^2+kN)\log N+kd(\log N)^2+d(\log N)^4(2\varrho\log_2N+1)^{2d}\Big)
\]
operations, plus the cost of factoring at most $\log_2N$ polynomials $P_i\in\mathbb F_{p_i}[x_1,\dots,x_d,t]$ with $\deg_tP_i=n_i$ and $\deg_{x_h}P_i\leq2n_i\varrho$, where $\sum_in_i\leq\log_2N$.
\end{theorem}

By~\cite{Lenstra1985}, a polynomial over a finite field $\mathbb{F}_{p_i}$ can be factored in time polynomial in its degrees and in the field size $p_i$. For each polynomial $P_i$, we have $\deg_t P_i=n_i$ and $\deg_{x_h}P_i\leq 2n_i\varrho$ for every $h$. Hence, for fixed $d$, its dense representation has at most
\[
(n_i+1)(2n_i\varrho+1)^d
\]
coefficients. Since $n_i\leq\log_2 N$ and $p_i\leq N$, factoring all at most $\log_2 N$ polynomials takes time polynomial in $N$ and $\varrho$, for fixed $d$. Consequently, including the factorization step, {\tt IsST} runs in time polynomial in $N$, $k$, and $\varrho$ for every fixed dimension $d$.

\section{Conclusions}\label{sec:conclusions}
 
We have shown that strong transitivity is decidable for group cellular automata over every finite group and in every dimension. GCAs over non-abelian groups are never strongly transitive, and for GCAs over abelian groups the decision reduces, via a decomposition by power subgroups which preserves strong transitivity, to linear CAs over groups of the form $(\mathbb Z/p\mathbb Z)^n$. For these, strong transitivity is characterized by the absence of irreducible factors of the characteristic polynomial with an invertible constant term. Pontryagin duality is the key tool in this characterization: it turns strong transitivity into the triviality of the eventual image of the dual endomorphism, a purely algebraic condition. 
For every fixed dimension $d$, the resulting decision procedure runs in time polynomial in the group size $N=|G|$, the number $k$ of neighbours, and  the neighbourhood radius $\varrho$, including the factorization step.

Several questions remain open.
\begin{itemize}
\item The cost of our procedure is dominated, for large neighborhoods, by the computation and the factorization of the characteristic polynomials. It would be interesting to know whether the existence of an irreducible factor with an invertible constant term can be decided without computing a complete factorization.
\item Strong transitivity of GCAs could be studied on group shifts, \ie closed shift-invariant subgroups of $G^\zzd$, instead of the full shift, as well as for CAs on more general amenable groups than $\zzd$.
\item Finally, it would be interesting to extend our results to CAs over finite monoids, and to other dynamical properties related to strong transitivity, such as openness and the various notions of expansivity.
\end{itemize}


\bibliographystyle{abbrv}
\bibliography{bib}

\end{document}